# Sub 100 nW volatile nano-metal-oxide memristor as synaptic-like encoder of neuronal spikes

Isha Gupta[a,*], Alexantrou Serb[a], Ali Khiat[a], Ralf Zeitler[b], Stefano Vassanelli[c], Themistoklis Prodromakis[a].

[a]Department of Electronics and Computer Science, Faculty of Physical Science and Engineering, University of Southampton, University Road, SO17 1BJ, Southampton, United Kingdom.

[b]Max Planck Institute for Intelligent Systems, Heisenbergstr,3,70569 Stuttgart, Germany.

[c]Department of Biomedical Sciences, University of Padova, Via Francesco Marzolo 3, Padova 35131, Italy.

**Advanced neural interfaces mediate a bio-electronic link between the nervous system and microelectronic devices, bearing great potential as innovative therapy for various diseases. Spikes from a large number of neurons are recorded leading to creation of big data that require on-line processing under most stringent conditions, such as minimal power dissipation and on-chip space occupancy. Here, we present a new concept where the inherent volatile properties of a nano-scale memristive device are used to detect and compress information on neural spikes as recorded by a multi-electrode array. Simultaneously, and similarly to a biological synapse, information on spike amplitude and frequency is transduced in metastable resistive state transitions of the device, which is inherently capable of self-resetting and of continuous encoding of spiking activity. Furthermore, operating the memristor in a very high resistive state range reduces its average in-operando power dissipation to less than 100 nW, demonstrating the potential to build highly scalable, yet energy-efficient on-node processors for advanced neural interfaces.**

Reverse engineering the human brain and decoding the underlying information processes of biological systems requires integrated efforts from researchers with different scientific backgrounds[1]. Towards enabling this vision, advances in neural recording techniques[2,3,4,5,6] target the reliable acquisition of electrophysiological data from multiple neurons *in-vitro* and *in-vivo*. This has impacted our understanding of information processing by brain microcircuits[7] and brought new prospects for novel therapies based on adaptive neural stimulation[8]. To date, state-of-art implementations can simultaneously record *in-vivo*[9] from up to thousand sites and from up to 30k[10] sites *in-vitro* using Complementary Metal Oxide Semiconductor (CMOS) based High Density Microelectrode Arrays (HDMEA's). Such advances in micro-sensors technology have been paralleled by considerable progress in neural processing microsystems[11,12] which are capable of detecting neural spiking activity on-node[13,14]. The relevant spike-detected information is then transmitted off-line wirelessly and techniques such as the Template Matching System (TMS) or Principle Component Analysis (PCA) [15] are used off-line for spike-sorting[16]. These methods, by mapping the recorded neural activity to the source active neurons, offer insights in neural coding principles[17] and support novel neuroprosthetic applications[18,19,20,8]. Thus, further advances in the fast developing field of implantable neural interfaces[21] are hampered by key bottlenecks in the processing of neuronal spikes including: a) computational power required to process the ever increasing volume of neural signals (Gb/s range presently) on-node and in real-time [22,23,24,25], b) bandwidth[26] and, c) scalability.

Recently, we proposed a new spike-detection approach[27] based on metal-oxide resistive switching memory devices, also known as memristors[28,29,30]. Fundamentally, memristive devices undergo non-volatile resistive state transitions as a function of the integral of the input voltage, thus behaving as thresholded input integrators[31]. Taking advantage of this property, we demonstrated that $TiO_x$-based memristive devices can be employed for spike-detection[27], as extracellular neural spikes recorded from retinal ganglion cells[32,33,34] were encoded in gradual, non-volatile resistive state transitions, whereas the sub-threshold events (i.e. noise) were naturally filtered-off[27]. This property makes these devices suitable as noise-suppressing integrating sensors and are thus termed as 'Memristive Integrating Sensors (MIS).' Non-volatility, however, was strongly limiting detection performance, as after saturation of the resistive state the devices, it failed to register any significant neural activity[35]. Consequently, performance was optimised by manual operation through frequent resets to the initial devices' resistive state,[36] which however impacts negatively on the overall power consumption.

In this work we advance on our previous findings by exploiting an often overlooked crucial property of memristive devices that is 'volatility'[37,38,39,40,41]. This approach recalls the way of operation of biological synapses that translate spiking frequency in gradual changes of postsynaptic conductance subject to a continuous self-resetting process[38,42]. When used in the volatile operating region, memristive devices exhibit metastable memory state transitions following which they inherently relax to their initial resistive state range. We demonstrate that volatility enables for naturally encoding spiking events into transient resistive changes. The self-resetting mechanism of our devices, ensures that these operate far from their resistive saturation region, which overall enhances the attained spike-detection accuracy. Moreover, we particularly exploit the fact that volatile phenomena are more pronounced at higher resistive states[39,41] reducing the overall power dissipation to less than 100 nW, setting a new state-of-art in spike detectors[25].

**Operation of nanoscale memristive devices as volatile cells**

Nanoscale TiO$_x$ metal-oxide memristive devices with metal-insulator-metal architecture, as shown in Fig.1a, were fabricated on a Si/SiO$_2$ substrate as detailed in the Methods section (Device Fabrication). A preliminary characterisation of the devices is needed to identify appropriate operation conditions in the volatile region[43]. A custom hardware infrastructure[44] was used for the electrical characterisation (see Methods, Hardware Infrastructure and Supplementary Figure 1). In brief, the devices are subjected to a high voltage stress until there is a sudden non-volatile change in their resistive state, known as electroforming[45,27]. The devices can then be operated in either non-volatile[27] (see Methods, Device Electrical Characterisation in non-volatile regime) or volatile manner depending upon the polarity of the voltage stress applied during the electroforming procedure and the strength of stimuli[38]. Importantly, the devices feature inherent threshold levels, below which there is no change in the resistive state and above which resistive state transitions are observed.

When operated in the volatile region our prototype devices undergo metastable resistive transitions within a high resistive state range and are capable of inherently re-attaining their approximate initial resistive state. To enable this study, we specifically developed a volatility characterisation algorithm[43] (see Methods, Device Electrical Characterisation in Volatile regime and Supplementary Figure 3) that applies a series of progressively more invasive voltage pulses and then monitors the resistive state of the device under test (DUT) (Fig. 1b and c) and its retention timescale. The module uses a standard two mean t-test to analyse the resistive state decay over time, terminating when the devices relax back to an equilibrium condition. We note that the algorithm makes no assumption on what the equilibrium resistive state should be. The equilibrium condition eventually corresponding to a non-volatile residual change of resistance of the DUT is determined through a retention test, which is implemented throughout a user-defined time window (for instance 60s in Fig. 1c). Notably, as shown in Fig.1 b and c, the operating resistive state region of the DUT was approximately 700 kΩ – 1.4 MΩ, using negative as the dominant stimulus polarity with 1 µs pulses (see Supplementary Figure 4 for 100 µs pulse width).

The output of the volatility module is estimates of resistive changes between the steady state and measurements taken immediately before and after the applied voltage stimulus (see Supplementary Fig.3b). This results in discrimination between non-volatile and volatile resistive changes for a given voltage stimulus as exemplified in Fig. 1d, where a -1.8V threshold voltage marks the transition of the DUT to a prevalent volatility state. The resistive state changes in the sub-threshold region are considered as insignificant and are mainly attributed to the background fluctuations caused due to the measurement noise. The range of identified inherent threshold voltages for the employed $TiO_x$-based memristive devices varied from approximately -0.6V to -2.5V (see Supplementary Figure 5). In conclusion, through the volatility module we determined the range in which the devices could be safely operated in the volatile region and roughly estimated the relaxation times to equilibrium.

The estimated volatility parameters were subsequently used when pre-processing the neural recordings in the spike-encoder platform. The platform's schematic is illustrated in Fig. 1e and an overall picture of the spike-encoding system including the front-end system is presented in Supplementary Figure 6. The neural recordings were obtained from slices of mid-peripheral rabbit retinal ganglion cells placed above an external front-end CMOS based MEA (see Methods, 'Front-end neural recording system CMOS Multi Electrode Array'). Importantly, in this work there was no modification performed on the front-end system, which was kept completely external to our experimental platform. Neural recordings arrive from the front-end as voltage-time data series in the range of approximately ±0.5V. For our experiments blocks of neural recordings containing approximately 63k samples recorded at a sampling rate of 12.2 kHz were used. In the first stage ('i'- Fig. 1b) the signals are pre-processed using a suitable Gain ($G$) and Offset ($V_{off}$) value. The neural signals are amplified such that the spikes – but not the noise – are above the volatility threshold of the DUT.

Pre-processed neural recordings are then passed through the memristive devices in batches ('ii'- Fig. 1e) and the resistive state of the DUT is read periodically in real-time. In this work, we follow a standardised signal processing flow: For each thousand data points batch, the resistive state of the DUT is recorded at the beginning, then every 300 points and finally at the end of the batch. This leads to segmentation of the neural data into smaller bins. Pairs of consecutive measurements in each batch are used to estimate the resistive state changes, whilst the pair of resistive state measurements taken at the end of each batch and the beginning of the following is used to estimate noise or reference data values (see Methods, **'**Neural recording biasing strategy' and Supplementary Figure 7). Finally, resistive state changes compressed in this way are processed offline and compared to noise ('iii'-Fig. 1e), with significant changes being recognized as spikes.

**Memristive devices as volatile spike-detectors**

Employing $TiO_x$ memristive devices as volatile spike-detectors requires pre-processing any neural recording input to match the volatile operating region of devices as determined using the volatility characterisation algorithm. To illustrate the concept, we chose a neural recording that contains a dense spiking pattern as illustrated in Fig. 2a. The operational parameters i.e. gain and offset values used for pre-processing the input neural recording were for this case fixed at 3.2 and 0, respectively. On biasing the target device with the neural recording the intrinsic reset capability of the DUT can be clearly noted in Fig. 2b. For instance, the initial resistive state of the DUT is approximately 350 kΩ following which the device demonstrates metastable resistive state transitions towards a low resistive state in response to supra-threshold spiking events. In the case where no subsequent supra-threshold events occur, the state of the device relaxes back towards the initial device state, demonstrating an inherent reset. The close-ups of the Fig. 2a, b in window 2.5s – 3.5s are presented in Supplementary Figure 8 to illustrate this point more clearly.

The number and approximate timing of spikes are estimated after post-processing of the resistive state measurements obtained using the standard schematic described in the methods section ('Neural recording biasing strategy'). As shown in Fig. 2e, resistive state change ($\Delta R/R_0$) in each bin is plotted as a function of highest voltage magnitude in each bin. The resistive state change magnitude distribution of noise measurements is used for identifying significant resistive state modulation that corresponds to spiking events. The inset of Fig. 2e represents the histogram for the noise measurements indicating an excessive inclination towards the positive polarity. Since the dominant stimulus polarity is negative and as a result neuronal activity-induced resistive changes are in the negative direction, the noise measurements in the positive direction are completely discarded and only the measurements in the negative direction are used to estimate meaningful noise band boundaries (see Methods, Neural signal processing and Supplementary Figure 9). This filters out the intrinsic reset transitions, which occur exclusively in the positive direction. Thus, noise band boundaries are estimated using only negative polarity noise measurements with a 4σ method (assuming Gaussian distribution), as indicated by the horizontal dashed line in Fig. 2e. Everything outside this band in the negative region is considered as significant modulation corresponding to a spiking event whilst all events registered within this band are disregarded as these do not correspond to state modulations due to spiking events. Following this methodology, the total number of spikes detected by our system, as shown in Fig. 2c, is equal to 67.

The performance of our memristor spike-detector is benchmarked against the established state-of-the-art template matching system[36] (TMS, see Supplementary Figure 10). As depicted in Fig. 2d, the total number of spikes determined by the TMS is equal to 78. In Fig. 2e, approximately -1.5V represents the inherent threshold voltage of the DUT. The negative quadrant (green) represents the spikes detected by our platform i.e. sum total of True Positives (TP) and False Positives (FP) whilst the positive quadrant indicates the False Negatives (FN) and True Negatives (TN). These quantification parameters are used to evaluate the sensitivity of our system through the true positive rate (TPR) and false positive rate (FPR) of detection (Methods section, 'Neural signal processing'). Assuming TMS to be a perfect spike detector the two values are estimated to be 74.35% and 5.14% respectively. Both values represent a significant improvement in performance compare to when the devices are operated in the non-volatile region. For example, for the same neural recording in Supplementary Figure 11, TPR and FPR after introducing optimised manual frequent resets is equal to 60% and 30% respectively.

The concept of volatile spike-detection via memristive devices was initially validated in Fig. 2 using a memristive device of dimensions 60 µm x 60 µm. The robustness of the devices for the same dimensions is illustrated in Supplementary Figure 12. A neural recording with significantly different spiking pattern in comparison to Fig. 2a was used to bias a memristive device. TPR and FPR in this case were equal to 65% and 0%, respectively. Besides, similar results from fifteen different devices are tabulated in Supplementary Table 1, where the highest TPR and FPR obtained is equal to 88.4% and 13% respectively.

**Nanoscale memristive devices as volatile spike-detectors**

Memristive device technologies offer huge advantages in terms of scalability and can be accommodated in Back-End-Of-Line (BEOL) of CMOS technologies, thus greatly benefiting future implantable neuroprosthetic platforms. We further support the presented approach with downscaled memristive devices of 200nm x 200nm dimensions. The devices were prototyped using the fabrication procedure described in the methods section (see Device Fabrication). In this case, we employed the same neural recording used previously, as shown in Fig. 2a and similarly benchmarked the obtained results against the TMS. For this case the neural recording was amplified using $G$ and $V_{off}$ values of 2.6 and -0.6 respectively, as illustrated in Fig. 3a. The transient response of the target device resistance is illustrated in Fig. 3b along with the corresponding neural recording time series. Compared to the previous case, this device was operated at a higher resistive state region of 1-1.5 MΩ. The spikes detected from both this and the TMS system was calculated to be 78, as shown in Fig. 3c and d respectively. The response of the target device can be more closely scrutinized in Fig. 3e and f, which illustrates the neural recording and resistive state response of the DUT during the 4-5sec window in Fig. 3a and b respectively. Marker 'x' is used to indicate the resistive state measurements, following the standard schematic as described in the methods section and detailed in Supplementary Figure 7. Between each pair of measurements the system is blind to the behaviour of the devices, however the bin size is a user-defined design parameter. Sampling rate (detection accuracy) and power consumption can be traded against each other.

The memristive devices undergo a resistive drop in response to the supra-threshold events. For instance, in the first bin shown in Fig. 3f, the state of the device drops from approximately 1.3 MΩ to 1.15 MΩ. The asterisk '*' in Figs. 3g, h further confirms concurrence between our system and the TMS that is true for the majority of instances. More specifically, the two systems agree for 13 over 17 instances detected, as shown in Fig. 3g. The symbol 'Φ' indicates an instance of mismatch between the two systems. Interestingly, at this instance our system detects a neural event that closely resembles a spike, while the TMS fails. On further careful examination of the neural recording it was observed that the TMS also fails to detect apparent spiking events at approximately 1.1s, 1.6s, 1.9s, 2.2s, 2.5s, 2.7s, 3.3s, 3.6s, 3.9s, some of which are detected by our system (see Supplementary Figure 13). On the other side, in Fig. 3g our system fails to detect spikes occurring at 4.55s. From these observations it can be safely concluded that, although we assume the TMS to be a perfect spike-detector for our experiments, in practical operation this is not the case. Benchmarking of the detected spikes revealed a TPR and FPR of 70% and 13.7% respectively. Spike detection results carried out by thirteen different nano-devices with pre-processed neural recordings are reported in Supplementary Table 1.

A Receiver Operating Characteristic sensitivity curve, that is defined as the rate of TP vs FP, is illustrated in Fig. 4a for devices of different dimensions. The details for the quantification parameters are presented in the Supplementary Table 1. It can be seen that as the dimensions of the devices are reduced from 60µm x 60µm down to 200nm x 200nm the detection accuracy of the system is reduced. One way for optimising the performance of the downscaled devices is to tune the gain and offset parameters to an optimum level. We thus proceeded by performing additional tests with three different gain settings 2.2, 2.4 and 2.6 and a fixed offset setting at -0.6V, as illustrated in Fig. 4b. For every round of gain, the experiment was repeated for five times and the details of the quantification parameters are illustrated in Supplementary Table 2. The asterisk symbol, '*' represents the average of the quantification parameters for every round of gain. Thus, an improvement in the gain resulted into an increase in rate of TP from 23.1% to 46.7%, and these numbers can be plausibly further improved in the future by estimating the optimum bin size for practical implementations of our volatile spike-detector. Moreover, exploring different stacks of memristive devices with engineered volatile characteristics might also be investigated for optimising performance.

**Discussion**

A major prerequisite in the implementation of fully implantable neural interfaces is to develop electronic platforms capable of executing low-power, on-site signal processing of the acquired neural activity. In this work, we experimentally demonstrated that metal-oxide memristive nanodevices can be operated in their volatile region as scalable spike-detection elements. The optimised performance is primarily due to the intrinsic resetting capability of such elements that eliminates the need for manually resetting the device to avoid its saturation. The major advantage of the presented work lies in the prospects for dramatically reducing the power consumption per channel of implantable neural spike-detection circuits. Consider the example illustrated in Supplementary Figure 11: for a neural recording consisting of approximately 63k data-points the target device is manually reset eleven times using a pulse of positive polarity of 100 µs pulse width. Realistically assuming +3V as the operating voltage and 10kΩ as the resistive state of the device, the amount of power dissipated including the reset operation by this system will be approximately equal to 3 mW per channel. The detailed power estimation methodology is presented in Supplementary Note 1. The reset operation consumes about 250 nJ of energy and 11 such resets for one neural recording would consume 2.7 µJ of energy which will be conserved in the volatile region.

Additional power savings also stem from the fact that the volatile regime exploited in our case occurs at relatively high resistive state regions (e.g., consider the electrical characterisation results of nano-devices in Fig. 1c and d). Assuming the operating resistive state and pulse width of the DUT to be 1MΩ and 1 µs respectively, the amount of energy dissipated per channel can be estimated using the standard batch processing schematic. The read operation with standard read out voltage of +0.2V will be equal to 0.2 pJ (0.04 pJ multiplied by 5 reads per batch) and the write operation at 3V will cost 9 nJ (9pJ multiplied by 1000 samples per batch). The average power dissipated per channel at 12.2 kHz as the sampling frequency can thus be estimated to be approximately 100 nW (see Supplementary Note 2). Voltage-time trade-off[46] can further assist in reducing the power dissipation by one order of magnitude, for instance, operating the same with 100 ns pulse widths will further reduce the power dissipated to 10 nW per channel per device. Our measured results are already significantly lower than current state-of-the-art spike detectors projected at approximately 700 nW[25]. Naturally, a full system application would include other power overheads required by the memristor read-out and biasing circuitry, which are not considered in the present work as such circuitry will be cited in the periphery and would be shared (multiplexed) by each memristive device.

One can also argue that the usability of the memristive devices could be limited due to the recurrently observed device-to-device variation. Nonetheless, our approach exploits normalised changes in each bin instead of absolute changes, therefore minimising any performance compromises due to device variability. Besides, the introduced volatility characterisation module leads to an automatic en-masse characterisation of memristive devices and determines the safe region for operation of devices in the volatile region. Clearly, if higher and more invasive voltages were applied, the devices may switch to the non-volatile region and change their baseline operating region[43]. The focal point for building on the present work is to push the scaling limits for the employed devices towards deep-submicron arrays[47], along with optimised spike-detection capability. We note that the use of alternative or engineered materials as the devices' active cores could allow tuning the volatile characteristics of devices[40] and in turn the self-reset achieved by the spike-detector. On the other hand, these parameters will also crucially depend upon the specific application under study, accounting for the sensitivity required for a particular application.

We further note that as the resistive state modulation encountered in our prototype devices is related to the amplitude and polarity of the input neural signal, the richness of the signal is in principle preserved and can possibly open a new avenue for scalable and power efficient on-node spike-sorting[48]; a prerequisite for fulfilling the electroceuticals[49] and more broadly the bioelectronics vision. In summary, taking the cue from how biological synapses compress spiking information in post-synaptic conductance changes, we have demonstrated a novel concept for neural-spike detection and encoding using intrinsic volatile behaviour of nanoscale metal-oxide memristive devices. Our results prove that single nanoscale volatile devices are capable of identifying significant spiking activity in the input neural waveform in a highly power efficient manner, thus paving the way towards advanced neuroprostheses or applications such as bioelectronics medicines where the power dissipation remains as the major challenge.

**Method Summary**

**Device Fabrication:** All nano-devices exploited in this work, that is, Ti/Pt/TiO$_2$/TiN (5/10/10/40 nm) were fabricated as follow: 6-inch wafer was thermally oxidised to grow 200 nm SiO$_2$, which serves as an insulating layer. Then, direct write e-beam lithography method was adopted, using JEOL JBX 9300FS tool, to pattern the bottom electrodes (BEs) nanowires. Double layer resists were used to facilitate lift-off process of the BEs, which are constituted of 5 nm adhesive Ti layer and 10 nm Pt film. BEs were deposited using e-beam evaporation. Bottom access-electrodes (large features) were then defined via conventional photolithography patterning, e-beam evaporation of Ti/Au (5 nm/25 nm) and lift-off process. Access-electrodes connect the pads to the nanowires. To pattern the active layer, optical lithography, reactive sputtering and lift-off process were also used. 10 nm near-stoichiometric TiO$_2$ active layer was sputtered with Leybold Helios Pro XL Sputterer from a Ti metal target. Next, 40 nm thick TiN top electrode (TEs) nanowires and 25 nm thick Au top access-electrodes were obtained in a similar manner to BEs and to the bottom access-electrodes, respectively, with the TiN films deposited via a Leybold Helios Pro XL Sputterer.

On the other hand, micrometre sized devices exploited in this work were fabricated on top of Si/SiO$_2$ wafers, where the oxide layer was 200 nm thick. In each layer, three main patterning steps were processed, optical lithography, film deposition and lift-off process. For the first layer, 5 nm Ti and 10 nm Pt films were deposited via electron-beam evaporation to serve as BEs. In the second, magnetron reactive sputtering system was used to deposit the 25 nm near stoichiometric TiO$_2$ active film. In the final step, 10 nm Pt TEs were deposited using electron-beam evaporation system, as well. The final stack is constituted of Ti/Pt/TiO$_{2-x}$/Pt (5/10/25/10 nm).

**Hardware Infrastructure/Instrumentation:** The electrical characterisation in pulsing mode was implemented using in-house fabricated electronic hardware infrastructure based on mBED LPC1768 micro-controller board[50,44] (Supplementary Figure 1). This instrument is capable of addressing single, up to 32 devices, or crossbar arrays up to 1kb in size (32x32 devices). The instrument is used to directly test the devices on-wafer interfaced via a multi-channel probe card. The hardware platform is supported by custom-made Graphical User Interface (GUI) that permits device-by-device fully automated testing. The biasing schemes applied for read and write operations are the Vr (Fig.3 in reference[51]) and Vr/2 (Fig. 10b in reference[52]) schemes, which also helps in mitigating sneak-path effects.

**Device Electrical Characterisation in non-volatile regime:** In the non-volatile regime, the devices are electrically characterised in two stages. Initially, the devices undergo an electroforming step. This involves application of ramp of voltages on a pristine sample until a sudden, non-volatile change in the resistive state of the device is observed. This typically occurs in the range of +6-8V[27] and the device is then considered to be in low resistive state that is ON state. The resistive state of the device decreases from 10's of MΩ down to 10kΩ. Thereafter, a train of input programming pulses in alternating polarities is applied at a fixed duration of 100μs which leads to reversible resistive switching[44]. The resistive state of the DUT is read after each programming pulse at 0.5V. The device is switched to low resistive state (ON/SET state) and high resistive state (OFF/RESET state) with positive and negative polarity respectively after the applied stimulus exceeds the DUT's inherent threshold voltage. Importantly, when operated in their non-volatile region, devices gradually switch within a 2 kΩ to 15 kΩ range. The response of the DUT for one specific input stimulus can be fitted to a second order exponential function[27] indicating saturation of its resistive state due to continued operation. The devices are asymmetric and demonstrate slightly different threshold for distinct polarities. The inherent threshold voltage of the employed devices varies in the range of approximately ± 0.6-2.5V[27].

**Device Electrical Characterisation in volatile regime:** In the volatile-regime, the devices tend to undergo metastable state transitions following which they relax to an equilibrium resistive state in a finite time window[38]. The devices are electrically characterised in two stages. The electroforming stage is similar to the one described earlier (Methods, Device Electrical Characterisation in non-volatile regime). For prototype devices with μm scale active areas we used negative polarity pulses (-6V - -8V), while positive polarity pulses (+4V- +6V) were employed in the case of devices with nm scale (200nm x 200nm) active core areas. Importantly, in the volatile region devices operate in a rather high resistive state range of approximately 300 kΩ – 3MΩ. In the second stage, the devices are characterised using an algorithm developed specifically to evaluate the retention characteristics of our prototypes (Figure 1 b, c and Supplementary Figure 3). The module applies a series of progressively more invasive pulses and then estimates the resistive state of the DUT using the standard two mean t-test method over a fixed interval of time. In practice, the t-test captures the resistive state progressive decay of the DUT and the module terminates when an equilibrium state is achieved. Subsequently, the t-test is followed by a retention condition test. In this test, the equilibrium state of the device is checked for a user-defined period of time. At the end, the output of the algorithm determines the time elapsing to achieve the equilibrium condition and the voltage ranges under which the devices can be safely operated in the volatile region, as estimated by comparing non-volatile with volatile changes (Figure 1d and Supplementary Figure 4).

**Front-end neural recording system (CMOS Multi Electrode Array):** In this work, neural activity was recorded from slices of dissected mid-peripheral rabbit retinal ganglion cells using an extended-CMOS technology[34] (Supplementary Figure 6). The CMOS based multi-transistor array (MEA) consists of 128x128 sensor sites, which records the data at a sampling rate of 12.2 kHz and outputs a current time series containing approximately 63k samples. The sensor sites of the CMOS-MEA are insulated by an inert $TiO_2/ZrO_2$ layer and a thin metal layer beneath the oxide layer is connected to the gate of the field-effect transistor. The voltage changes due to the interfaced neural tissue/cells above the recording sites are used for modulating the source-drain current in the MOSFET. Trans-Impedance Amplifiers (TIA) fabricated on-chip convert the signal into voltage and amplify the signal from a 0.1 mV- 1 mV range up to 10-500 mV in range. This amplified signal is then used as an input for our platform. The CMOS MEA was kept external to the presented platform and in this work the CMOS MEA is termed as the 'front-end' system.

**Neural recording biasing strategy:** The input to our experimental platform is the neural signal recorded from the MEA-based CMOS system [34](see Methods, Front-end neural recording system and Supplementary Figure 6). Each neural recording obtained contains approximately 63k neural data points recorded at a sampling rate of 12.2 kHz. The neural recordings are in the range of ± 0.5 V. The obtained data is pre-processed using a software-based gain and offset stage. This suitably amplified neural trace is then used to bias individual memristive devices using the customised hardware (Figure 1e and Supplementary Figure 1) and the signal is fed to target devices in batches of 1000 data-points. In each processing batch, the resistive state of the device is assessed five times that is at the beginning of each batch, then after every 300 samples and finally at the end of each batch. Four consecutive measurements are thus obtained at $300^{th}$, $600^{th}$, $900^{th}$ and $1000^{th}$ data-point, and one measurement is made at the end of each batch and before the beginning of next batch without any neural data point in between. Our method transforms a batch of 1000 data-points in five bins and thus results into an overall data compression rate of 200. Resistive state changes can be extracted from the consecutive measurements (bins) whist the measurement uncertainty (N) can be estimated from the measurements made at the end of each batch and the beginning of the next batch. As a result, for a single neural recording (approximately 63k points) we obtain 316 resistive state measurements corresponding to 252 consecutive resistive state changes and 64 noise-level measurements. These resistive state changes help in estimating the threshold voltage of the target device and consequently differentiating the significant resistive state changes from the insignificant ones (see Supplementary Figure 7,9).

**Neural signal processing**: Estimation of the spikes detected by our platform involves post-processing of the resistive state measurements. The normalised resistive state changes in each bin are plotted as a function of the highest voltage magnitude in each bin (see Fig. 2e and Supplementary Figure 9). In the volatile region of operation, all the noise measurements made in the positive polarity are discarded. Only noise measurements in negative polarity are used for estimating the noise band boundaries. Absolute values of the noise measurements in the negative polarity are generated and then standard deviation (σ) is calculated. 4-σ method that is μ ± 2σ is used and the noise band boundary is set as shown by the horizontal red dashed line in Fig. 2e. All the resistive state changes outside this estimated boundary are considered to be significant whilst the resistive state changes falling within this band are disregarded. We argue that these resistive state modulations occur due to weak amplitude neural signals and cannot be differentiated from the noise measurements. On the other hand, if noise measurements in the positive direction are included, estimation of mean and standard deviation values will be significantly affected leading to higher probability of inclusion of noise. Moreover, for further clarity, the comparison of the noise band setting in the volatile region with the non-volatile region is presented in Supplementary Figure 9. The noise band and an ideal threshold voltage divides the resistive state measurements in four parts and the data is quantified as follows[36]: Outside the noise band and below

$V_{th}$: True Positives (TP); outside the noise band and above $V_{th}$: False Positives (FP); inside the noise band and below $V_{th}$: False Negatives (FN); inside the noise band and above $V_{th}$: True Negatives (TN). These results are benchmarked against state-of-the-art TMS and these parameters are then redefined. TP: indicates an agreement between the two systems for spike detected in a given bin. TN: means the two systems agree that there is no activity in a given bin, FP: our system detects an event whilst TMS system doesn't and FN: TMS system detects a spike whereas our system does not. Using these values, rate of TP (TPR) and FP (FPR) is determined using following equations:

$$Rate\ of\ TP = \frac{TP}{TP + FN}$$

$$Rate\ of\ TP = \frac{FP}{FP + TN}$$

**Data Availability:** The data that support the findings of this study are available from the corresponding author upon request, as detailed in http://www.nature.com/authors/policies/data/data-availablity-statements-data-citations.pdf.

**Supplementary Information** is available in the online version of the paper.

**Acknowledgements** We acknowledge the financial support of FP7 RAMP and EPSRC EP/K017829/1. Experimental procedures involving the use of animals were approved within the RAMP projects by Ethics Committee of the University of Padova and the Italian Ministry of Health (authorisation.447/2015-PR). All the experiments were conducted in accordance with the approved guidelines.

**Author contributions**: T.P., S.V. and I.G. conceived the experiments. A.K. optimised and fabricated the devices. I.G. and A.S. performed the electrical characterisation of the samples and developed the control instrumentation and software. R.Z. developed the front-end recording platform. All the authors contributed towards the analysis of the results and writing the manuscript.

**Author information** Reprints and permissions information is available at. The authors declare no competing financial interests. Correspondence and requests for materials should be addressed to I.G. (I.Gupta@soton.ac.uk).

**Figure 1** Device architecture and electrical characterisation of nanoscale TiO$_x$ memristive devices in the volatile region. (a) Schematic (right) and Scanning Electron Microscopic (SEM) image (left) of the employed 200nm x 200nm memristive device. (b,c) Electrical characterisation of memristive device exhibiting volatility (see Supplementary Figure 3 for a detailed description). In brief, an automatic volatility characterisation module[43] is utilised and the resistive state evolution of the DUT in response to the applied stimuli is illustrated in (c) The device operates in the 600 kΩ - 1.3 MΩ range. After a first electroforming stage (see Methods), the algorithm applies a series of progressively increasing input pulses (blue bars) of amplitude ($V_w$) and width ($T_w$) to the target device, following which its resistive state is monitored. In the example, the write pulses ($V_w$) were applied in the range of -0.2*V*- -4*V* in steps of 0.2V. $T_w$ was 1 μs. The module operates on a standard statistical two mean t-test condition and terminates when the device reaches a steady state. Thus, equilibrium retention is verified by repeated readings over 60s (marked as pink rectangles), following which a new stimulus is delivered. (d) Determination of the operating voltage range of the DUT in volatile region. For every step of input stimulus applied, volatile (green circles) and residual non-volatile (red circles) resistive state changes were measured. The grey band indicates the voltage region for ensuring operation of the DUT in volatile conditions, with approximately -1.8*V* being the inherent threshold voltage of the DUT ($V_{th}$). (e) Schematic for the implementation of our memristor-based spike-detection platform[27]. The green dashed box indicates the external front-end CMOS - based MEA from which the neural recordings are obtained. The spike-detection platform consists of three stages: (i) the recorded neural data is pre-processed using a gain (*G*) and offset ($V_{off}$) stage; (ii) the processed neural recordings are then used to bias the target memristive device using the hardware infrastructure and the resistive state of the device is read

periodically; (iii) the compressed resistive state measurements are processed offline and the significant resistive state changes are filtered-off from the insignificant ones. Significant resistive state changes are registered as spiking events.

**Figure 2** Spike-detection via volatile metal-oxide memristive device and benchmarking of the results against the TMS. (a) Neural recording used for biasing the DUT. Gain ($G$) and offset ($V_{off}$) values in this experiment were fixed at 3.2 and 0, respectively. The pink band indicates the inherent threshold voltage of the DUT ($V_{th}$) which is approximately equal to -1.5V. (b) Resistive state evolution of the DUT with time in response to the neural recording in (a). Time intervals where our spike-detector identifies events are indicated in grey. (c) total number of spikes detected by our system and (d) via the TMS. (e) Normalised changes in the resistive state of the device ($\Delta R/R_0$) in each bin are plotted as a function of highest voltage magnitude in each bin. The resistive state measurements in the yellow eclipse represents the noise measurements made at the end of each batch and beginning of the next batch with no neural feed. The inset represents the histogram for the noise measurements. The red dashed line on the horizontal axis represents the boundaries of the noise band estimated using the 4 σ method that is μ ± 2σ whilst the green band indicates the significant resistive state changes as detected by our system. True Positives (TP), False Positives (FP), True Negatives (TN) and False Negatives (FN). On benchmarking these values against the TMS, the rate of TP (TPR) and FP (FPR) are estimated.

**Figure 3** Scalability of the TiO$_x$ devices to the nanoscale dimensions and operation of nano-devices as volatile spike detectors. (a) The employed neural recording pre-processed using gain and offset value of 2.6 and -0.6 respectively. The pink band indicates the inherent threshold voltage of the DUT ($V_{th}$). (b) Resistive state changes in the target DUT in response to the neural data in (a). The active core area of the nano-devices used for this experiment is 200nm x 200nm. The threshold voltage of the DUT is -1.3*V*. (c), (d) Total number of spikes detected by our system and the TMS system is equal to 78 for both the systems. The black bins indicate the spike positions. (e) and (f) are close-ups for the neural recording and the resistive state evolution shaded grey in (a) and (b) respectively. 'X' cross mark in red indicates the positions where the resistive state measurements are taken. Time intervals where an event is detected are represented in grey. (g) spikes detected by our system and (h) the TMS system. The asterisk '*' indicates the positions where the two systems agree and 'Φ' symbol discusses a specific case of discrepancy between the two systems.

**Figure 4** Comparison of Receiver Operating Characteristics (ROC) that is rate of true positives vs rate of false positives for memristors of micro and nano dimensions. (a) The blue and red colour indicates the ROC curves for 60µm x 60µm and 200nm x 200 nm devices respectively. Symbol psi 'Ψ' and circle in blue colour indicates different neural recordings with different spiking pattern used for the DUT. (b) Optimisation of spike detection capability for one of the 200nm x 200nm device. Blue, red and green colour indicates the three different gain parameters i.e. 2.2, 2.4 and 2.6 respectively chosen for the experiment where the offset was kept constant at -0.6. For every gain the experiment was repeated five times. Asterisk symbol '*' indicates the average of the quantification parameters for each round of gain.

# Figure 1

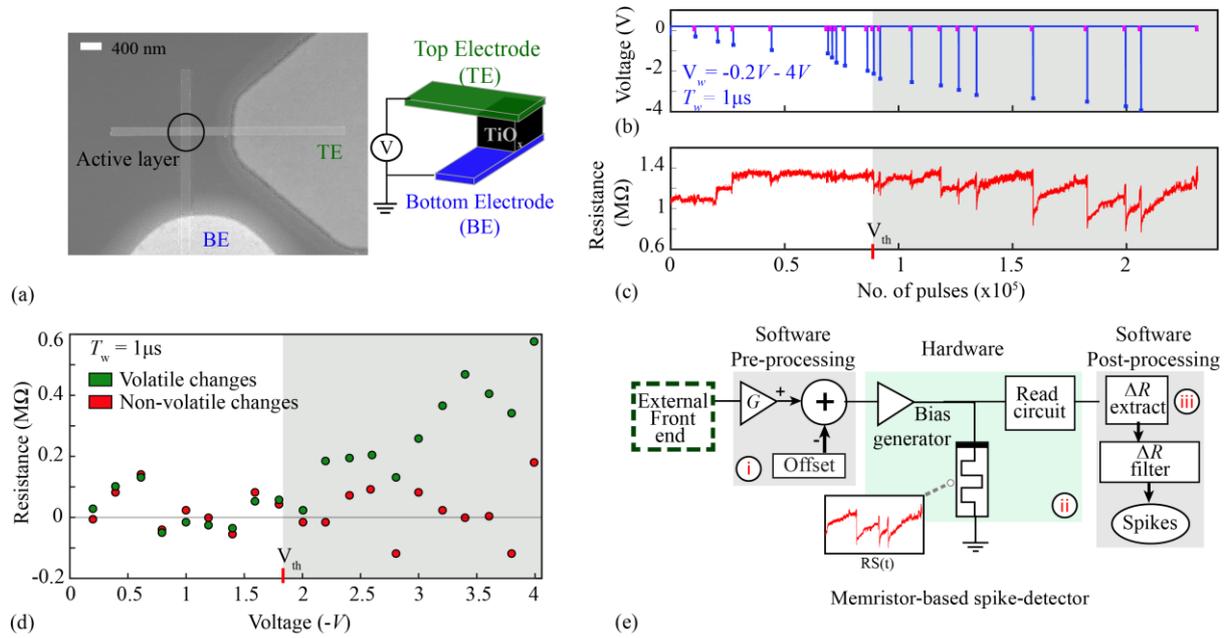

# Figure 2

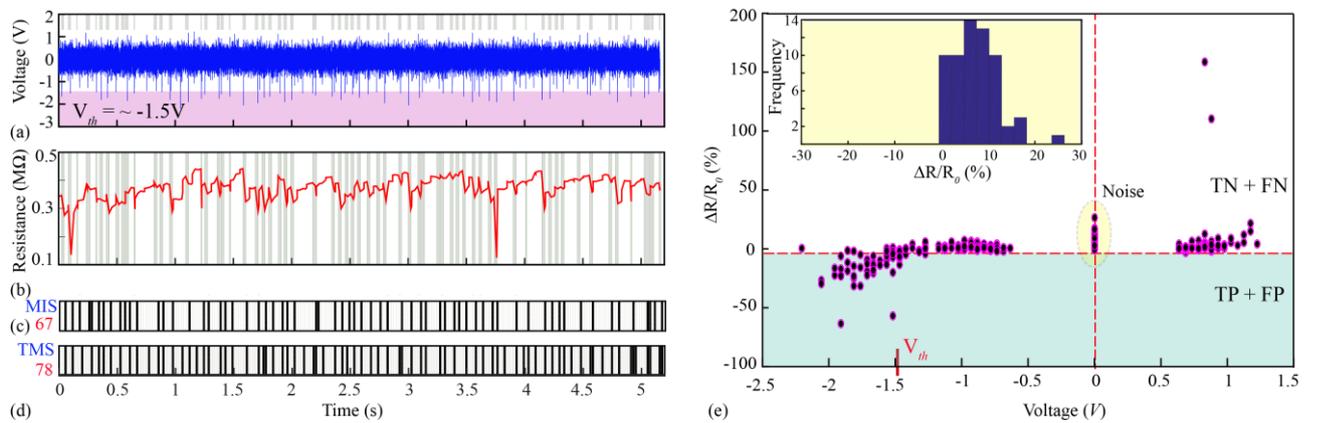

# Figure 3

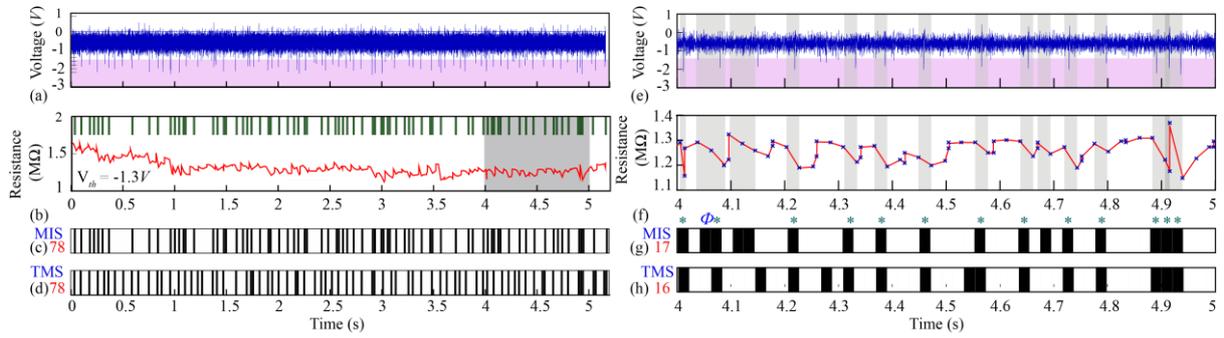

# Figure 4

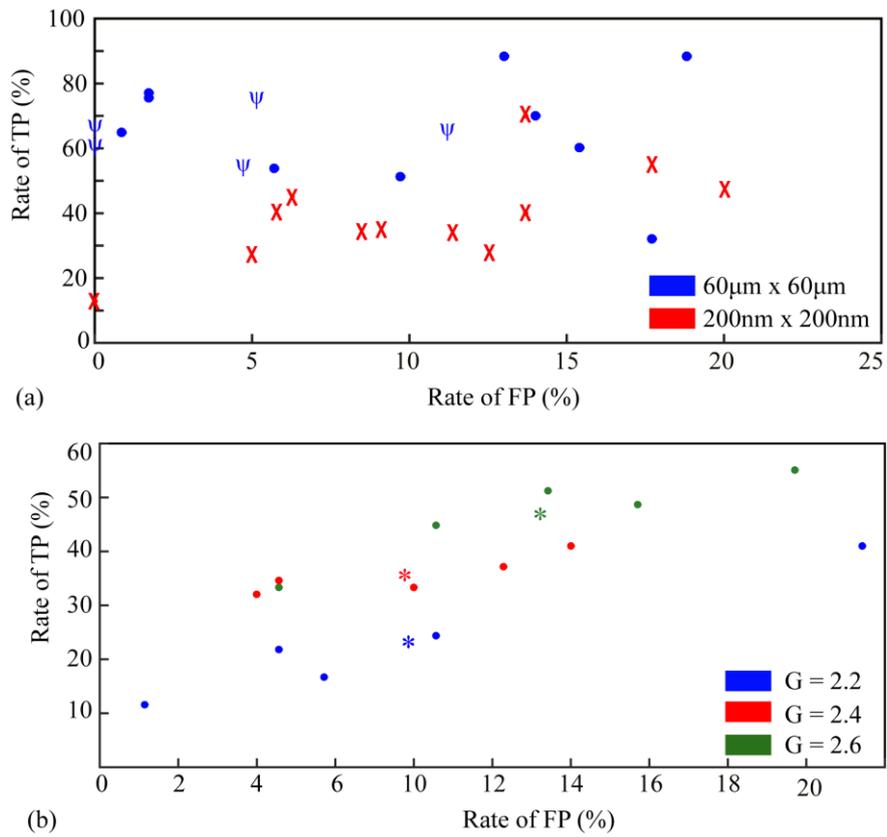

# SUPPLEMENTARY INFORMATION

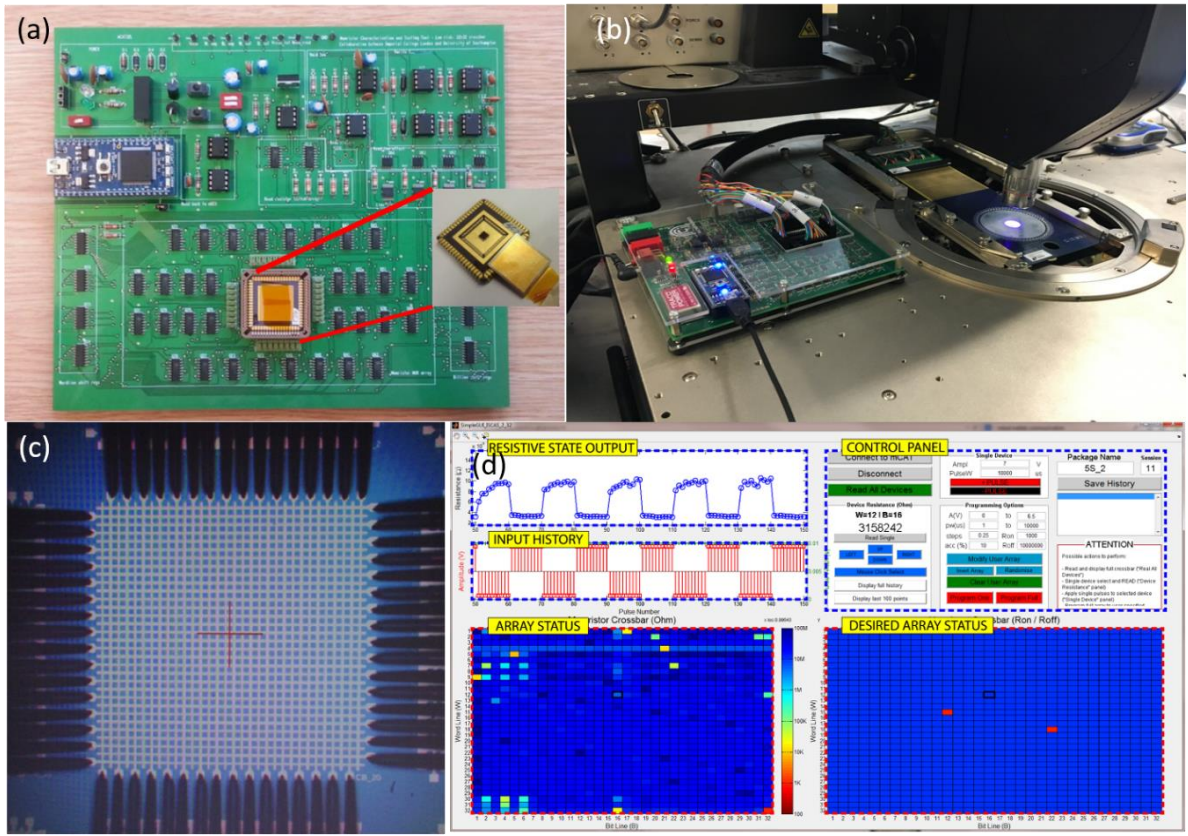

**Supplementary Figure 1** Hardware infrastructure used in order to implement the proposed biasing protocol in the MIS platform. (a) Illustration of the custom-made hardware developed in order to facilitate en-masse electrical characterisation of the fabricated memristive devices[1,2]. The red lines indicate the position of inserting the wire bonded packaged memristive devices. (b) On-wafer testing of the memristive devices using a probe card used as a communication channel between the mBED board and the wafer. At one point of time approximately 1000 devices fabricated in crossbar architecture or 32 devices in stand-alone configuration can be interfaced using the developed testing technology. (c) Illustration of the probe card needles touching down on one of the crossbar configurations. (d) Graphic user interface (GUI) supporting the hardware infrastructure. GUI can be used for programming the biasing protocols applied to the memristive devices and the status of the devices can then be tracked in real-time.

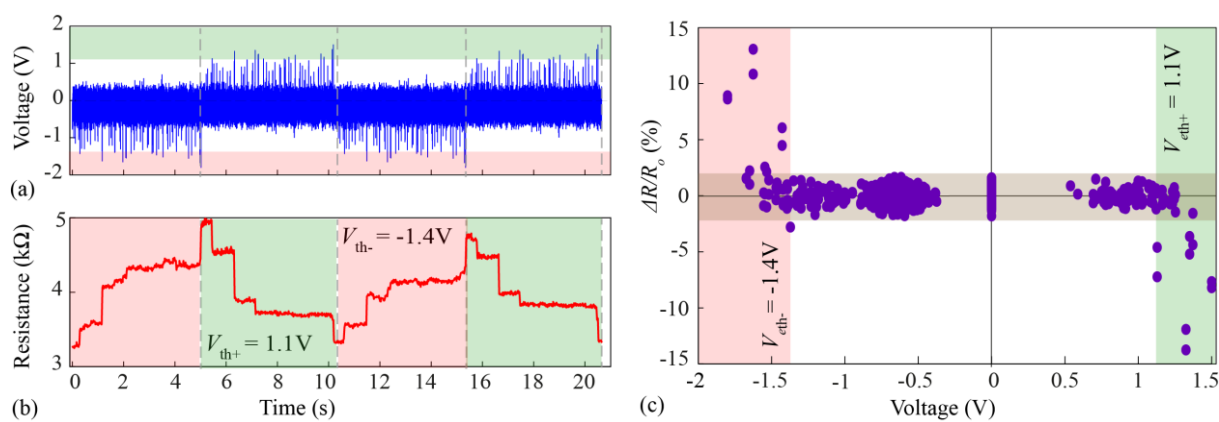

**Supplementary Figure 2**[3] Operation of TiO$_x$ devices in non-volatile region[3]. (a) A single neural recording is concatenated four times alternatively in opposite polarity. The selected neural recording was processed in MATLAB to invert the polarity. The green and pink band in the figure indicates the extracted threshold of the device-under-test (V$_{eth-}$). The G and V$_{off}$ value for this specific neural recording was set to 2.8 and 0, respectively. (b) Response of the device-under-test in response to the neural recording in (a). In response to the negative (positive) supra-threshold events there is gradual increase (decrease) in the resistive state of the device. The same phenomenon can be seen over the illustrated four cycles. Notably, the supra-threshold events are encoded in the resistive state changes of the devices. The threshold for the two polarities is slightly different indicating inherent asymmetry in the devices which is typical of this device family. (c) The resistive state changes in each bin are plotted as a function of the highest voltage magnitude in each bin. On x = 0 axis, the agglomerated changes are indicative of the noise measurements made at the end of each batch. These measurements are used to set the noise band marked by the grey band. Anything falling in this band is considered to be 'insignificant' and everything outside this band is considered to be a 'significant' change and is estimated as the spike estimated by the MIS platform.

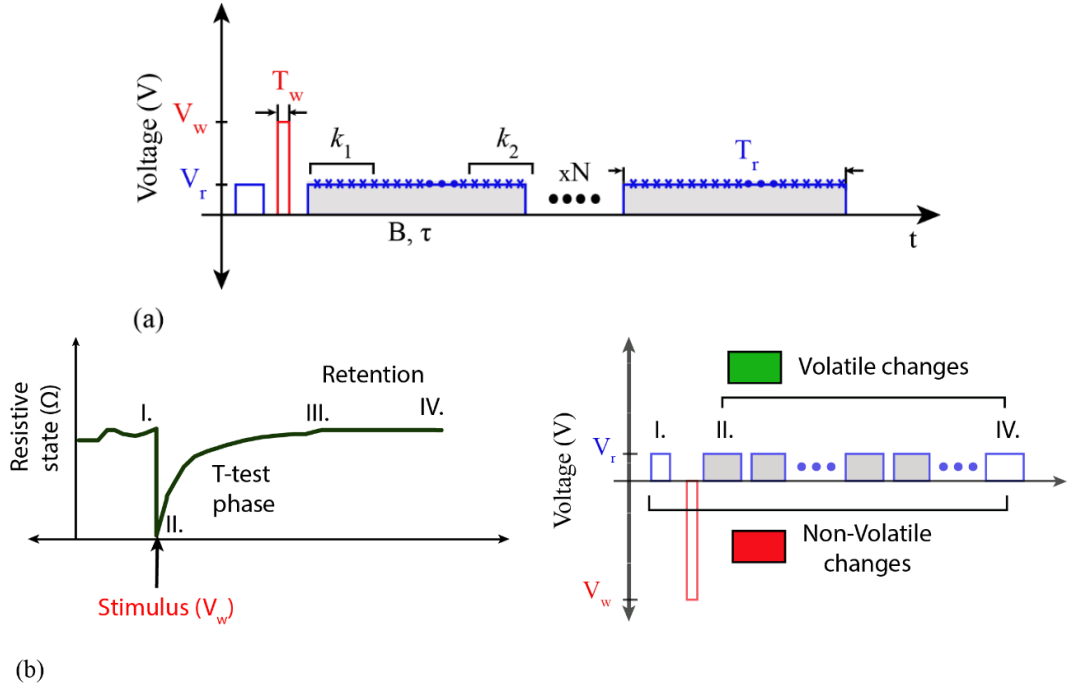

(a)

(b)

**Supplementary Figure 3** Description of the Volatility Characterisation Algorithm[4]. The algorithm is applied in two stages. In the first stage, a series of progressively more invasive voltage pulses are applied and resistive state decay over time is monitored. In the second stage, the equilibrium condition is checked using the retention condition for 60 seconds. (a) Schematic for volatility module. The schematic describes the first stage for the volatility module. A programming pulse with amplitude and duration ($V_w$ and $T_w$) is applied on the device-under-test. A read-voltage ($V_r$) of $0.2V$ (blue bars) is applied and the resistive state of the device is monitored using the standard two mean t-test in batches ($B$) with $n$ measurements, where $n$ is user-defined according to the equation 1. To estimate the output of t-test in each batch, the mean ($\mu$) and the standard deviation ($\sigma$) of the first and last $k$-values are estimated. If the estimated value is less than the set threshold voltage (set as '1' in this work), the device is assumed to have been relaxed to its equilibrium state and the algorithm is terminated otherwise the next batch is applied and the same procedure is repeated until the device reaches a steady-state. '$\tau$' describes the overall time of the t-test. (b) **Left Panel**: Four major points for the assessment of the resistive state of the devices in the volatile module. 'I' and 'II' is the measurement of the resistive state of the device before and after the application of the user-defined input stimulus. 'III' marks the termination of the t-test stage whilst 'IV' indicates the termination of the retention stage. **Right panel:** The estimation of the volatile (difference between the point IV and II) and non-volatile (difference between points IV and I) changes. The difference of volatile from non-volatile changes helps in determining the voltage range in which the devices can be safely operated in volatile region.

Equation 1:

$$t = \frac{\mu_1 - \mu_2}{\sqrt{\frac{\sigma_1^2}{k_1} + \frac{\sigma_2^2}{k_2}}}$$

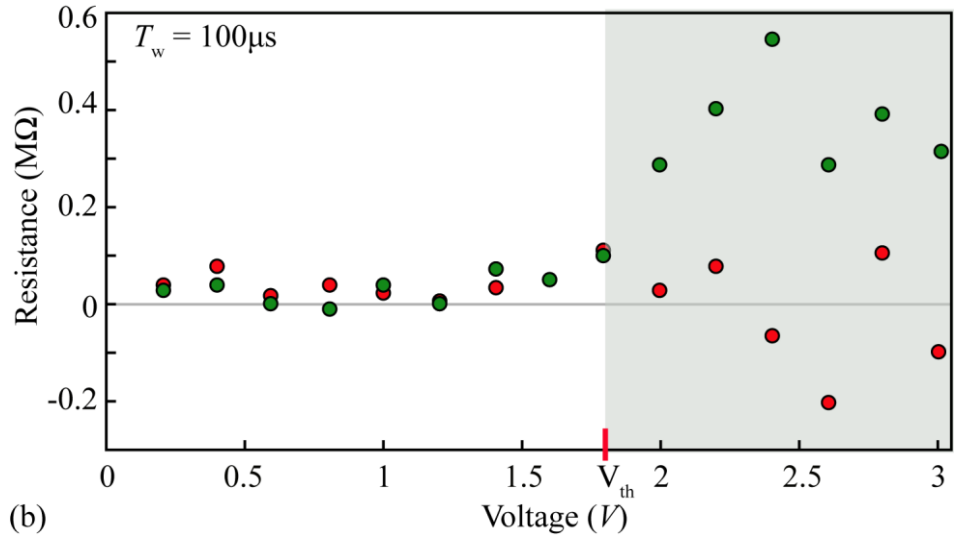

(b)

**Supplementary Figure 4** Volatility characterisation of the TiO$_x$ nano-devices (200 nm x 200 nm) using the volatility module described in Supplementary Figure 3[4]. Input stimulus was applied to the device-under-test in range of 0- -3V in steps of 0.2V. The pulse width, T$_W$, of the applied voltage was fixed, in this case, to 100 µs. Red and green circles indicate the non-volatile and volatile changes respectively. For the device-under-test the threshold of the device was found to be approximately -1.8V. The grey band indicates the region where the device can be operated in the volatile region.

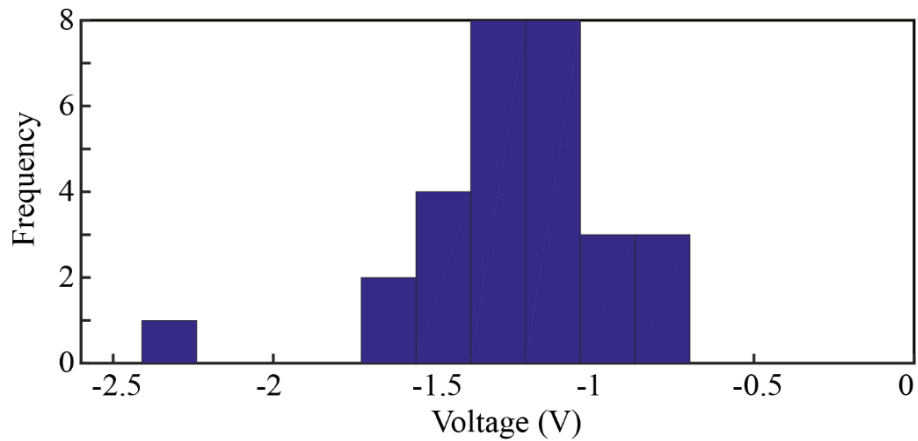

**Supplementary Figure 5** Range of inherent threshold voltages (V$_{th}$) for the memristive devices in volatile region as extracted from 29 different devices.

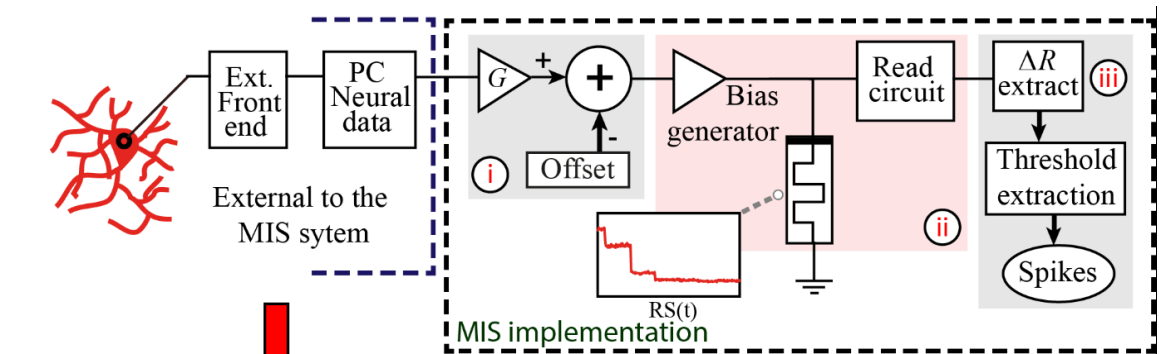

(a)

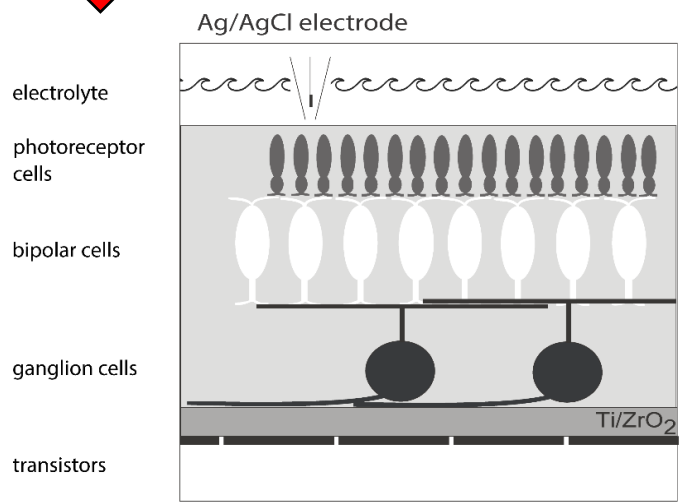

(b)

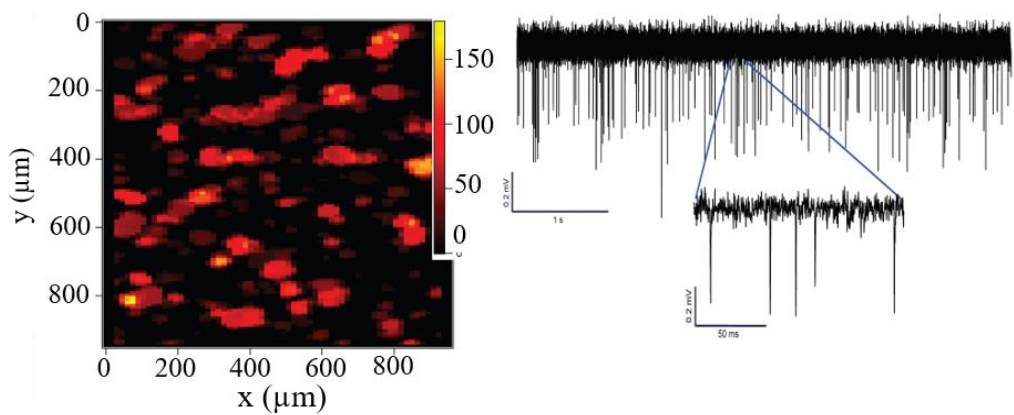

(c)

**Supplementary Figure 6** Overall block diagram for the memristor based spike-detector platform[3]. (a) Block diagram highlighting the fact that in this work the implementation of spike-detection platform indicated by the black dashed line is external to the Complementary Metal Oxide Semiconductor (CMOS) Multi Electrode Array (MEA) neural recording system[5,6,7]. The extracellular recordings obtained from the rabbit retinal ganglion cells using CMOS MEA that is termed as 'front-end' and is stored on to a PC. These recordings are acquired and processed using the spike-detection platform. As illustrated in (a) in the first step the recordings are suitably amplified using a 'gain ($G$)' and 'offset ($V_{off}$)' stage (i). The pre-processed recordings are then used to bias the memristive devices using the hardware infrastructure described in Supplementary Figure 1, where the resistive state of the device is read periodically (ii). The compressed read-out states are processed offline and the spikes detected by the spike-detection platform are estimated (iii). (b) Image of the retina/chip configuration in the employed CMOS MEA atop which the slices from the mid-peripheral of the rabbit retina are placed and measured directly. The transistors are separated from the cell layer using a thin Ti/ZrO$_2$ insulating layer. (c) (Left) Surface plot of the electrical activity from tissue slices placed atop the CMOS MEA. (Right) Blocks of neural activity (voltage-time series) in hundreds of mV range obtained from the front-end system.

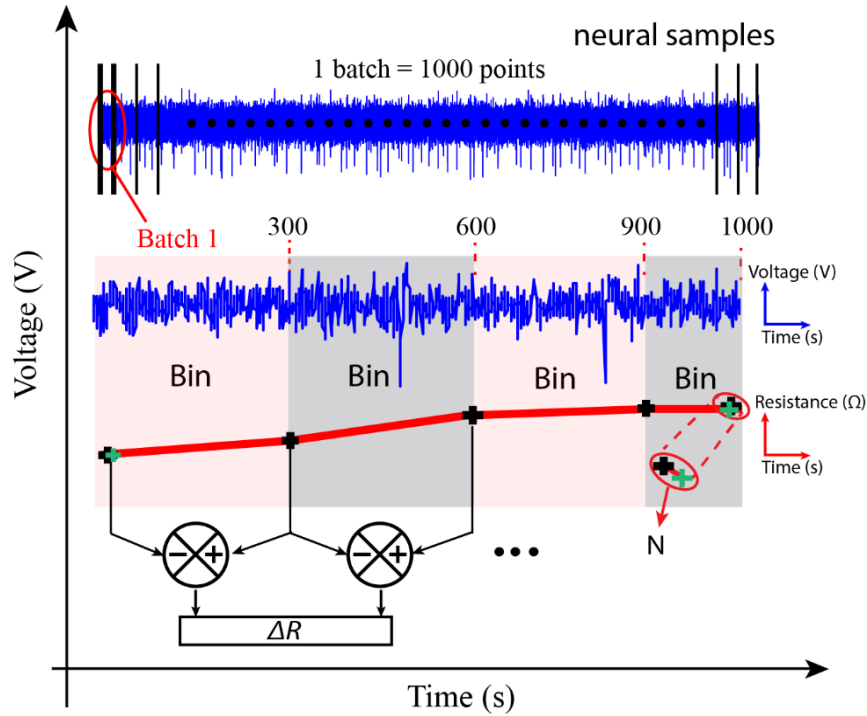

**Supplementary Figure 7** Signal processing methodology in the spike-detection platform[3]. Every block of neural recording obtained from the 'front-end' system is approximately 63k samples. The neural data after suitable amplification using the G and V$_{off}$ stage is fed to the memristive devices in batches of 1000 data points. In every single batch (1000 data points) the resistive state of the device is assessed five times in smaller bins, i.e. at the beginning of each batch and then after every three hundred points. For instance, in the first 1000 neural data points, the resistive state is read after 300, 600, 900 and 1000[th] point. These consecutive measurements are used to estimate the resistive state changes in each bin. Notably, the neural recording is paused during the reading of the resistive state of the device. One measurement is also taken at the end of each batch and at the beginning of the next one, with no neural data in between. These measurements are used to estimate the Noise ($N$) in the system. $N$ can arise both from measurement uncertainties of the employed devices and from the intrinsic noise of the measurement board. Hence, the described methodology transforms 63k neural data points into 316 resistive state measurements, containing noise estimates. Thus, data are compressed by a factor of 200. Thereafter, resistive state measurements are post-processed to estimate the resistive state changes. Significant resistive state changes (explained in Supplementary Figure 9) are estimated as spike count of the spike-detection platform.

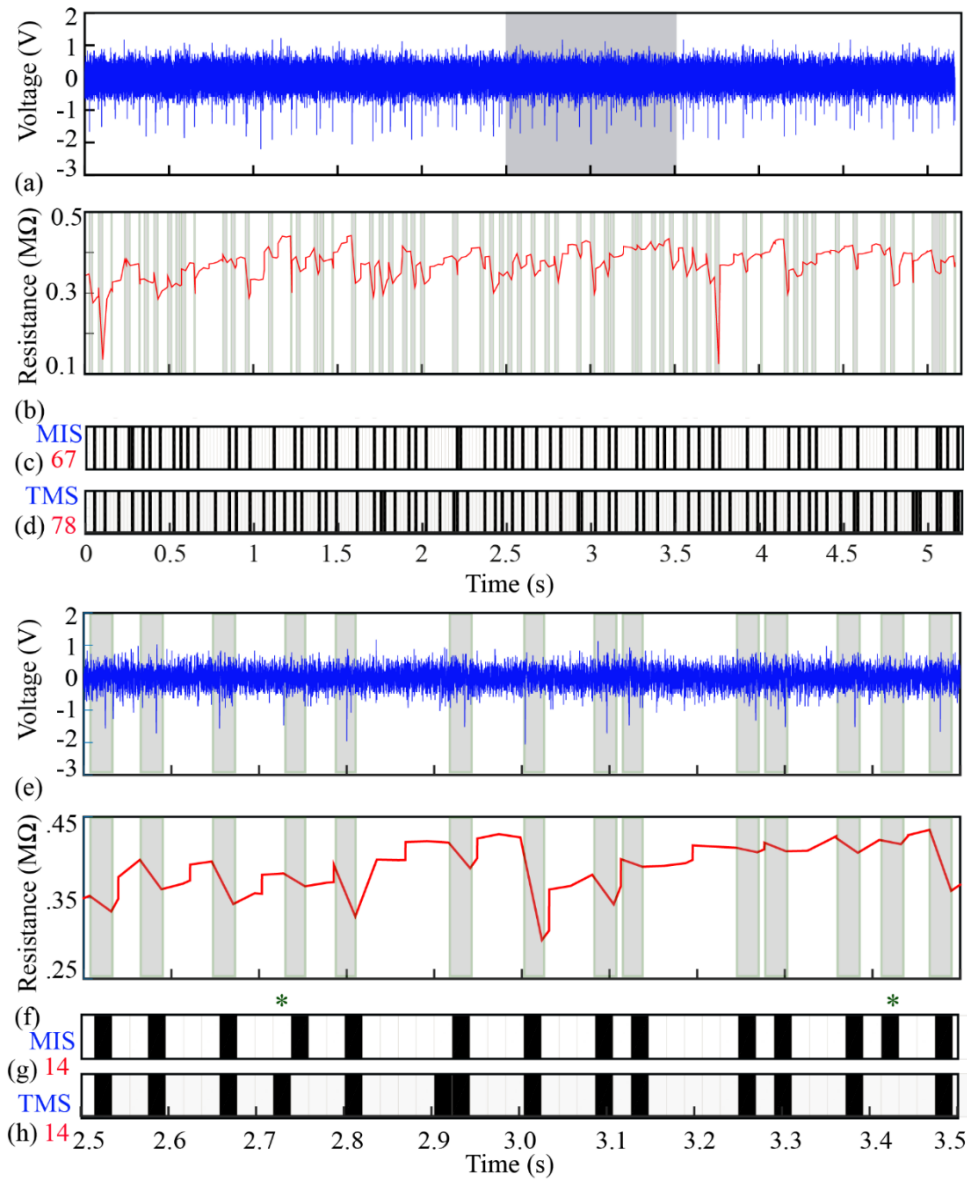

**Supplementary Figure 8** Biasing of 60 µm x 60 µm TiO$_x$ devices using a neural recording in the volatile region. (a) Neural recording used for the biasing of the device-under-test. (b) Resistive state evolution of the device-under-test with time. (c) and (d) Spikes detected by the MIS platform (67) and TMS (78) respectively. (e) and (f) Close-up of the neural recording employed in (a) and the resistive state evolution in (b) for time window 2.5s -3.5 s, respectively. (g) and (h) Spikes detected by the MIS and TMS system, respectively. In the close-up window the two systems agree for majority of instances except the point marked by '*'. The grey band indicates the bins with significant resistive state changes estimated as spikes in the output of the MIS platform. The quantification parameters for this specific recording are as follows: TP, FP, TN and FN are 58, 9, 166 and 20 respectively. The rate of TP (TPR) and FP (FPR) for this specific recording is estimated to be 74.3 and 5.14 respectively.

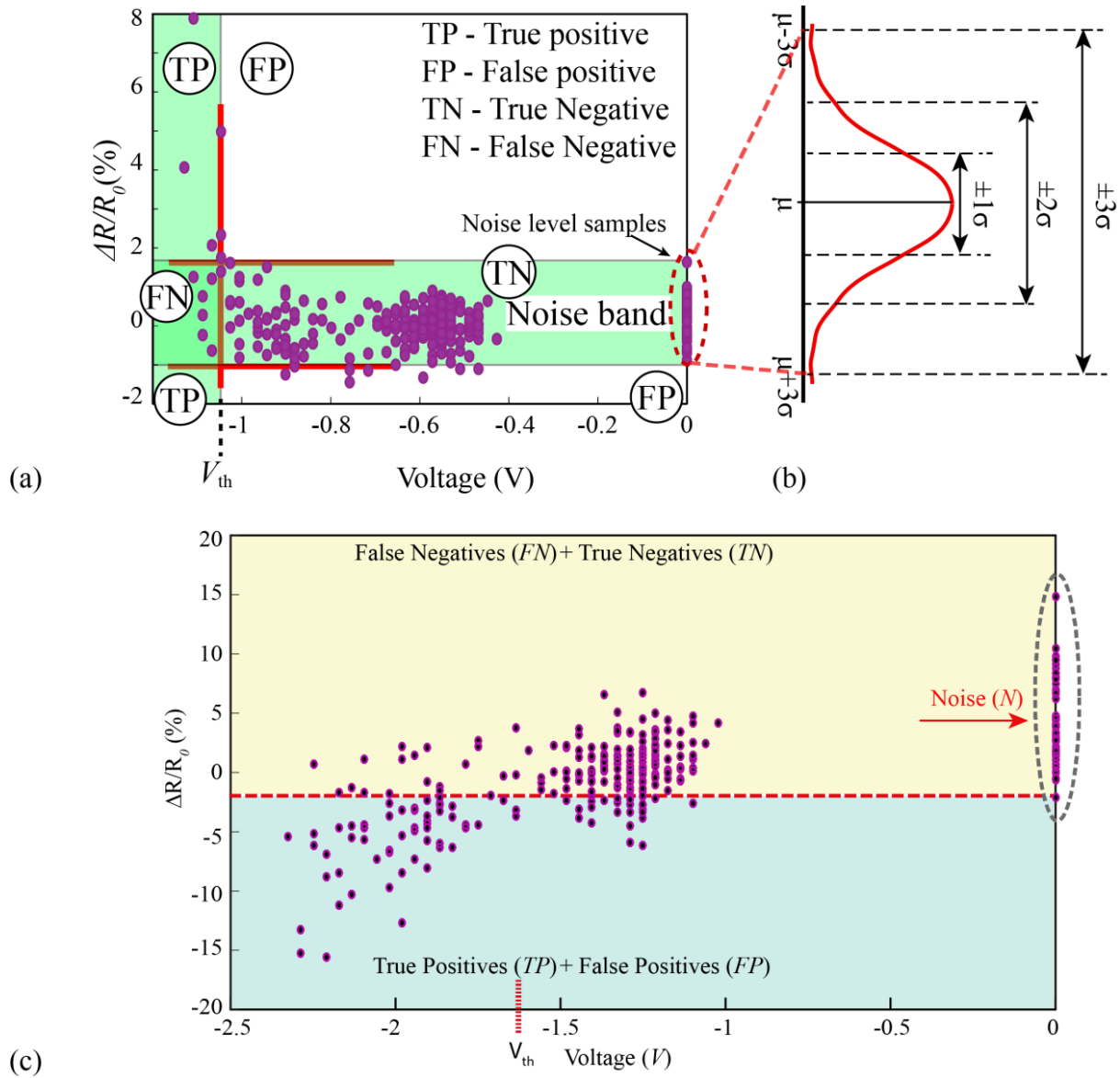

**Supplementary Figure 9** Comparison of the Noise band diagrams in the volatile and non-volatile region of operation[8]. (a) Noise-band diagram settings in the non-volatile region[8]. Normalised plot of resistive state changes ($\Delta R/R_0$) in each bin is plotted as a function of maximum voltage magnitude in each bin. The noise-band (horizontal band marked in green colour) is estimated using the noise band measurements made at the end of each batch and the beginning of next batch with no interceding neural data points (as explained in Supplementary Figure 7). (b) Assuming Gaussian distribution, mean and the standard deviation are estimated. 6 sigma method is chosen to set the boundaries of the noise-band. The estimated resistive state changes falling within this band are discarded as this cannot be differentiated from the noise band measurements. Everything outside this band is estimated to be a significant change and is estimated as a spike in the output of the spike-detection platform. The existence of the threshold voltage classifies the resistive state changes in four groups. In a simple threshold detector, everything above the threshold ($V_{th}$) is estimated as a spike and everything below is discarded as noise. However, because of the existence of noise-band in the spike-detection platform, the data is quantified as follows: Outside the noise band and below $V_{th}$: True Positives (TP), outside the noise band and above $V_{th}$: False Positives (FP), inside the noise band and below $V_{th}$: False Negatives (FN), inside the noise band and above $V_{th}$: True Negatives (TN).

The results are benchmarked against the established template matching system and the quantification parameters are redefined. TP (TN) indicates when the two system agree (disagree) for the presence of spike. FP indicates a spike detected by our platform and not by the TMS system. FN indicates a spike indicated by the TMS system and not spike-detection platform. Importantly, in this work we assume template matching system to be a perfect spike detector. The rate of TP (TPR) and FP (FPR) are estimated using the following equations:

$$Rate\ of\ TP = \frac{TP}{TP + FN}$$

$$Rate\ of\ TP = \frac{FP}{FP + TN}$$

(c) Noise-band diagram in the volatile region. For the experiments with neural recordings in the volatile region most of the significant spikes are present in the negative polarity. Noise measurements are marked using a grey dashed eclipse indicating the spread of measurements which notably is inclined in the positive polarity. During the operation of the device in the volatile region, the significant resistive state changes due to supra-threshold spikes are in the negative quadrant. This is due to the fact that the metastable resistive state transition is to a low resistive state following which the device intrinsically resets to a higher resistive state. Hence, for the noise band boundary settings in the volatile region, the noise band measurements in the positive polarity are discarded (that is switching of the device-under-test back to high resistive state) and only the measurements in the negative polarity are used. The boundary is set using the 4-sigma method which is indicated by the horizontal red dashed line dividing the resistive state changes in two quadrants i.e. True Positive + False Positives (spikes detected by our platform) and False Negatives + True Negatives. Again, the quantification and benchmarking of the results against the template matching method is carried out using the equations described in (b).

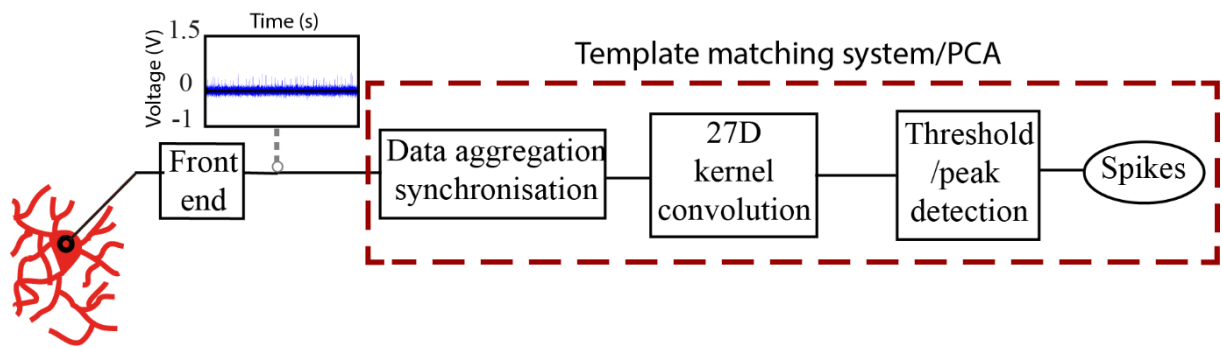

**Supplementary Figure 10** Schematic of the Template Matching System (TMS) used as a benchmarking standard for the proposed volatile spike-detection platform[3]. The same front-end system as described in Supplementary Figure 6 is used to acquire the neural data. The data in this system is processed through computationally heavy Principal Component Analysis (PCA) technique[6].

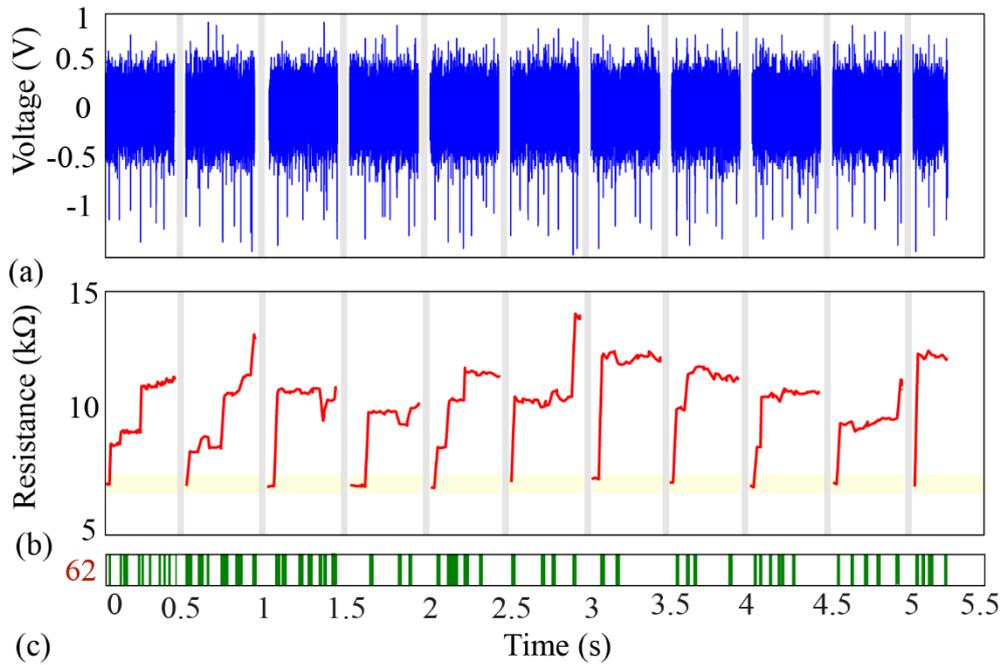

**Supplementary Figure 11** Manual frequenting of the memristive devices in the non-volatile region[8]. The continued operation of the device in a single polarity leads to the saturation of the resistive state of the device. As a mitigation strategy, the neural recording of approximately 63k data points is sliced into smaller sub-neural recordings with approximately 6k data points. Grey bands indicate the separation regions, as shown in (a). (b) Resistive state response of the DUT in correspondence to the sub-neural recordings with time. After every sub-neural recording the device is manually reset to its initial resistive state represented in yellow bands with a pulse of positive polarity of 100 µs. The initial resistive state of the device is in the region of 6-8 k$\Omega$ and the operation of the device is in the region of 6 k$\Omega$ (low resistive state) to 15 k$\Omega$ (high resistive state). (c) The total spike count detected is 62 represented as bins in green colour.

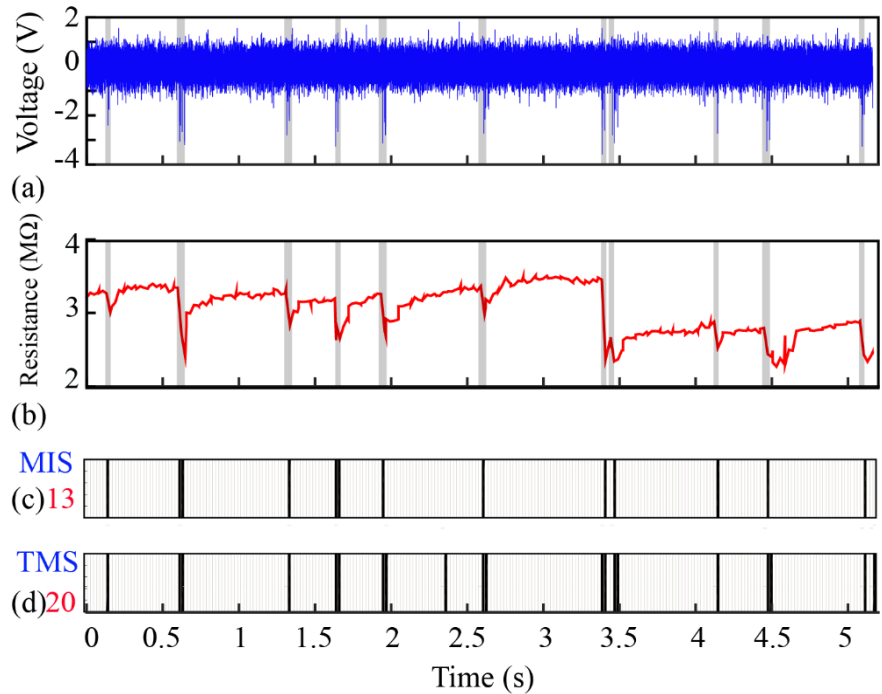

**Supplementary Figure 12** Robustness of TiO$_x$ devices. Biasing of 60 µm x 60 µm devices using a different neural recording in comparison to the one presented in Supplementary Figure 8 in the volatile region. (a) Neural recording used for the biasing of the device-under-test. (b) Resistive state evolution of the device-under-test with time. (c) and (d) Spikes detected by our platform (13) and TMS (20) respectively. The grey band indicates the bins with significant resistive state changes estimated as spikes in the output of our platform. The quantification parameters for this neural recording are as follows: TP, FP, TN and FN are 13, 0, 233 and 7 respectively. The rate of TP (TPR) and FP (FPR) is found to be 65 and 0 respectively.

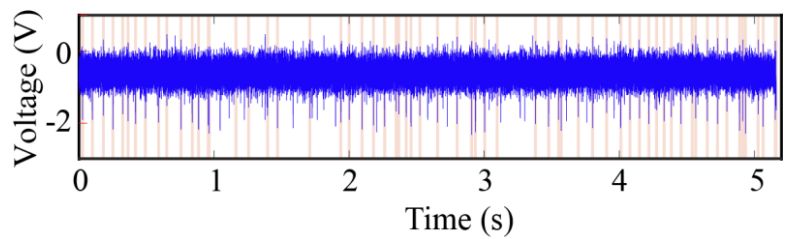

**Supplementary Figure 13** Spikes detected by the Template Matching system. The red bands depict the spikes detected. Notably, few instances where the TMS clearly fails to detect neural events are at approximately 1.1s, 1.6s, 1.9s, 2.2s, 2.5s, 2.7s, 3.3s, 3.6s, 3.9s.

| Device Dimensions | $V_{th-}$ (V) | Gain | Offset | VMS | TMS | TP | FP | TN | FN | Rate of TP (%) | Rate of FP (%) |
|---|---|---|---|---|---|---|---|---|---|---|---|
| 60 x 60 µm² | -1.4 | 3.2 | 0 | 67 | 78 | 58 | 9 | 166 | 20 | 74.35 | 5.14 |
| 60 x 60 µm² | -1.4 | 2.6 | 0 | 39 | 20 | 13 | 26 | 207 | 7 | 65 | 11.15 |
| 60 x 60 µm² | -1.17 | 2.4 | -0.6 | 22 | 20 | 11 | 11 | 222 | 9 | 55 | 4.7 |
| 60 x 60 µm² | -1.7 | 2.4 | -0.4 | 15 | 20 | 13 | 2 | 231 | 7 | 65 | 0.85 |
| 60 x 60 µm² | -1.1 | 2 | 0 | 15 | 20 | 13 | 2 | 231 | 7 | 65 | 0.85 |
| 60 x 60 µm² | -0.7 | 2.2 | 0 | 74 | 78 | 47 | 27 | 148 | 31 | 60.25 | 15.4 |
| 60 x 60 µm² | -0.7 | 2.2 | -0.2 | 52 | 78 | 42 | 10 | 165 | 36 | 53.8 | 5.71 |
| 60 x 60 µm² | -1.34 | 2.2 | 0 | 12 | 20 | 12 | 0 | 233 | 8 | 60 | 0 |
| 60 x 60 µm² | -1 | 4.4 | 0 | 56 | 78 | 25 | 31 | 144 | 53 | 32 | 17.7 |
| 60 x 60 µm² | -1 | 4.4 | 0 | 57 | 78 | 40 | 17 | 158 | 38 | 51.3 | 9.7 |
| 60 x 60 µm² | -1 | 4.8 | 0 | 76 | 78 | 53 | 23 | 152 | 25 | 70 | 14 |
| 60 x 60 µm² | -1.2 | 4.8 | -0.2 | 102 | 78 | 69 | 33 | 142 | 9 | 88.46 | 18.8 |
| 60 x 60 µm² | -1.2 | 2.9 | 0 | 62 | 78 | 59 | 3 | 172 | 19 | 75.6 | 1.71 |
| 60 x 60 µm² | -1.2 | 2.6 | 0 | 63 | 78 | 60 | 3 | 172 | 18 | 77 | 1.71 |
| 60 x 60 µm² | -2.41 | 4.4 | 0 | 13 | 20 | 13 | 0 | 233 | 7 | 65 | 0 |
| 60 x 60 µm² | -0.7 | 2.6 | -0.4 | 92 | 78 | 69 | 23 | 152 | 9 | 88.4 | 13 |
| 200x200 nm² | -1.1 | 2.8 | -0.4 | 54 | 78 | 31 | 24 | 151 | 47 | 40 | 13.7 |
| 200x200 nm² | -1.25 | 2.6 | -0.6 | 46 | 78 | 35 | 11 | 164 | 43 | 44.8 | 6.28 |
| 200x200 nm² | -1.3 | 2.6 | -0.6 | 30 | 78 | 21 | 9 | 166 | 57 | 27 | 5 |
| 200x200 nm² | -1.63 | 2.8 | -0.4 | 10 | 78 | 10 | 0 | 175 | 68 | 12.8 | 0 |
| 200x200 nm² | -1.25 | 2.6 | -0.6 | 42 | 78 | 32 | 10 | 165 | 46 | 41 | 5.74 |
| 200x200 nm² | -1.25 | 2.6 | -0.6 | 78 | 78 | 54 | 24 | 151 | 24 | 70 | 13.7 |
| 200x200 nm² | -1.3 | Max = 0 | Min= -2.2 | 44 | 78 | 22 | 22 | 153 | 56 | 28.2 | 12.5 |
| 200x200 nm² | -1.13 | 2.6 | -0.6 | 47 | 78 | 27 | 20 | 155 | 51 | 34.6 | 11.4 |
| 200x200 nm² | -1.2 | 3 | -0.6 | 47 | 78 | 27 | 20 | 155 | 51 | 34.6 | 11.4 |
| 200x200 nm² | -1.3 | 3 | -0.6 | 42 | 78 | 27 | 15 | 160 | 51 | 34.6 | 8.5 |
| **200x200 nm²** | **-1.3** | **2.6** | **-0.6** | **74** | **78** | **43** | **31** | **144** | **35** | **55.12** | **17.71** |
| 200x200 nm² | -1.4 | 2.6 | -0.6 | 72 | 78 | 37 | 35 | 140 | 41 | 47.43 | 20 |
| 200x200 nm² | -1.45 | 2.6 | -0.6 | 43 | 78 | 27 | 16 | 159 | 51 | 34.61 | 9.14 |

**Supplementary Table 1** Robustness of TiO$_x$ memristive devices. For this experiment, devices with different dimensions i.e. 60 µm x 60 µm and 200 nm x 200 nm and different neural recordings with significantly different spiking pattern were used. For the pre-processing of the neural recording, the operational parameters that is *G* and V*$_{off}$* were varied. The quantification parameters are indicated in the table with the estimated rate of true positives and false positives. VMS:Spikes detected by our platform, TMS: Template matching system, TP: True Positives, FP: False Positives, TN: True Negatives, FN: False Negatives.

| Device Dimensions | Gain | Offset | VMS | TMS | TP | FP | TN | FN | Rate of TP (%) | Rate of FP (%) |
|---|---|---|---|---|---|---|---|---|---|---|
| 200x200 nm$^2$ | 2.2 | -0.6 | 66 | 78 | 32 | 34 | 141 | 46 | 41.02 | 19.42 |
| 200x200 nm$^2$ | 2.2 | -0.6 | 25 | 78 | 17 | 8 | 167 | 61 | 21.79 | 4.57 |
| 200x200 nm$^2$ | 2.2 | -0.6 | 23 | 78 | 13 | 10 | 165 | 65 | 16.67 | 5.71 |
| 200x200 nm$^2$ | 2.2 | -0.6 | 34 | 78 | 19 | 15 | 160 | 59 | 24.35 | 8.57 |
| 200x200 nm$^2$ | 2.2 | -0.6 | 11 | 78 | 9 | 2 | 173 | 69 | 11.53 | 1.14 |
| | | | | | | | | | 23.07 | 7.88 |
| 200x200 nm$^2$ | 2.4 | -0.6 | 32 | 78 | 25 | 7 | 168 | 53 | 32.05 | 4 |
| 200x200 nm$^2$ | 2.4 | -0.6 | 47 | 78 | 29 | 18 | 157 | 49 | 37.18 | 10.28 |
| 200x200 nm$^2$ | 2.4 | -0.6 | 40 | 78 | 26 | 14 | 161 | 52 | 33.34 | 8 |
| 200x200 nm$^2$ | 2.4 | -0.6 | 53 | 78 | 32 | 21 | 154 | 46 | 41.02 | 12 |
| 200x200 nm$^2$ | 2.4 | -0.6 | 35 | 78 | 27 | 8 | 167 | 51 | 34.61 | 4.57 |
| | | | | | | | | | 35.64 | 7.77 |
| 200x200 nm$^2$ | 2.6 | -0.6 | 62 | 78 | 38 | 24 | 151 | 40 | 48.71 | 13.71 |
| 200x200 nm$^2$ | 2.6 | -0.6 | 50 | 78 | 35 | 15 | 160 | 43 | 44.87 | 8.57 |
| 200x200 nm$^2$ | 2.6 | -0.6 | 60 | 78 | 40 | 20 | 155 | 38 | 51.28 | 11.42 |
| 200x200 nm$^2$ | 2.6 | -0.6 | 34 | 78 | 26 | 8 | 167 | 52 | 33.34 | 4.57 |
| 200x200 nm$^2$ | 2.6 | -0.6 | 74 | 78 | 43 | 31 | 144 | 35 | 55.128 | 17.71 |
| | | | | | | | | | 46.66 | 11.19 |

**Supplementary Table 2** Optimisation of 200nm x 200nm TiO$_x$ memristive devices. Gain and offset parameters for one of the 200nm x 200nm device was optimised. Three different values of gain i.e. 2.2, 2.4 and 2.6 were used with constant offset values i.e. -0.6. For each round of gain the experiment was repeated five times. The quantification parameters for each round when benchmarked against the state-of-the-art template matching system are indicated in the illustrated table. VMS: Spikes detected by our platform, TMS: Template matching system, TP: True Positives, FP: False Positives, TN: True Negatives, FN: False Negatives.

# SUPPLEMENTARY MATERIAL

**Supplementary Note 1:**

**Estimation of power dissipated per channel using manual frequent method in Supplementary Figure 11 (Non-volatile region).**

Assumptions:

1) The following calculations have been done assuming the resistive state of the device to be 10 kΩ and series resistance (compliance) to be 1 kΩ.
2) The read voltage of the device is +0.5V and the write voltage is conservatively assumed to be +5V.
3) The pulse width used for the experiment is 100 µs.
4) For every batch of thousand data points, the state of the device is read five times.
5) P: Power, R: Resistance, V: Voltage, E: Energy, t: time (employed pulse width).
6) Sampling frequency is 12.2kHz.

**Energy consumed for read operation**:

$P = V^2/R = (0.5)^2/(1+10)$ kΩ $= 0.023$ mW

$E = P \times t = 0.023$ mW x $100$ µs $= 2.3$ nJ

E dissipated for 5 read operations = 2.3 nJ x 5 = ==11.5 nJ==

**Energy consumed for write operation**:

$P = V^2/R = (5)^2/ (10)$ kΩ $= 2.5$ mW

$E = P \times t = 2.5$ mW x $100$ µs $= 250$ nJ

For 1000 samples, 1000x 250 nJ = ==250 µJ==

**Resetting operation using one single pulse**:

$E = 2.5$ mW x $100$ µs = ==250 nJ==

**Average Power Consumption** :

~ 3mW.

**Supplementary Note 2:**

**Estimation of power dissipated per channel in volatile MIS platform using Fig.1 b and c.**

Assumptions:

1) The following calculations have been done assuming the resistive state of the device to be 1MΩ and series resistance (compliance) to be 100 kΩ.
2) The read voltage of the device is +0.2V and the write voltage is conservatively assumed to be +3V.
3) The pulse width used for the experiment is 1 µs.
4) For every batch of thousand data points, the state of the device is read five times.
5) P: Power, R: Resistance, V: Voltage, E: Energy, t: time (employed pulse width).
6) Sampling frequency is 12.2kHz.

**Energy consumed for read operation**:

$P = V^2/R = (0.2)^2/(1\text{M}\Omega+100$ kΩ$) = 0.04$ uW

$E = P \times t = 0.04$ uW $\times$ 1 us $= 0.04$ pJ

E dissipated for 5 read operations $= 0.04$ pJ $\times$ 5 $=$ ==0.2 pJ==

**Energy consumed for write operation**:

$P = V^2/R = (3)^2/(1)$ M$\Omega$ $= 9$ uW

$E = P \times t = 9$uW $\times$ 1us $= 9$ pJ.

For 1000 samples (per batch), 1000$\times$ 9 pJ $=$ ==9 nJ.==

**Average Power Consumption**:

0.2 pJ+ 9 nJ/0.082 $=$ ==~100 nW.==

Importantly, if the same experiment is performed with 100 ns pulses, the power consumption would be approximately 10 nW per batch.

---------------------------------------------------------------------------------------------------------------------------------